# Radiative Relaxation Quantum Yields for Synthetic Eumelanin[‡]

Paul Meredith[*1] and Jennifer Riesz[1]

[1] Department of Physics & Centre for Biophotonics & Laser Science, University of Queensland, Brisbane, Queensland 4072, Australia

Dr. Paul Meredith, University of Queensland Department of Physics, St. Lucia, Brisbane, QLD4072, Australia

**Phone**: +61 7 3365 7050

**Fax:** +61 7 3365 1242

**e-mail**: meredith@physics.uq.edu.au

**Abbreviations:** DHI, 5,6-dihydroxyindole; DHICA, 5,6-dihydroxyindole, 2-carboxylic acid; Dopa, dihydroxyphenylalanine; PL, photoluminescence; PLE, photoluminescence excitation

**ABSTRACT**

We report absolute values for the radiative relaxation quantum yield of synthetic eumelanin as a function of excitation energy. These values were determined by correcting for pump beam attenuation and emission re-absorption in both eumelanin samples and fluorescein standards over a large range of concentrations. Our results confirm that eumelanins are capable of dissipating >99.9% of absorbed UV and visible radiation through non-radiative means. Furthermore, we have found that the radiative quantum yield of synthetic eumelanin is excitation energy dependent. This observation is supported by corrected emission spectra, which also show a clear dependence of both peak position and peak width upon excitation energy. Our findings indicate that photoluminescence emission in eumelanins is derived from ensembles of small chemically distinct oligomeric units which can be selectively pumped. This hypothesis lends support to the theory that the basic structural unit of eumelanin is oligomeric rather than hetero-polymeric.

## INTRODUCTION

The melanins are an important class of pigmentary macromolecules found throughout nature (1). Eumelanin is the predominant form in humans, and acts as the primary photoprotectant in our skin and eyes. Physiochemically, all melanins are broadband UV and visible light absorbers, and potent free radical scavengers and antioxidants (1,2). In direct contradiction with its photoprotective properties, eumelanin (along with pheomelanin – the less prevalent red-brown pigment also found in humans), is implicated in the cytotoxic chain of events that ultimately lead to melanoma skin cancer (3). For this reason, the photophysics, photochemistry and photobiology of melanins are subjects of intense scientific interest.

In a more general sense, the broader structure-property-function relationships that dictate the behavior of these important biological macromolecules are still poorly understood (4). In particular, major questions still remain concerning the basic structural unit (5). It is fairly well accepted that eumelanins are macromolecules of DHI and DHICA, and that pheomelanins are cysteinyl-dopa derivatives (6,1). However, it is still a matter of debate as to whether eumelanin (in particular) is actually a highly cross-linked extended heteropolymer, or composed of DHI/DHICA oligomers condensed into 4 or 5 oligomer nano-aggregates (7). This is an absolutely fundamental issue, and is the starting point for the construction of consistent structure-property-function relationships. The answer to this question also has profound implications for our understanding of the condensed phase properties of melanins. In 1974, McGinness, Corry and Proctor showed that a pellet of dopamelanin could be made to behave as an amorphous electrical switch (8). They postulated that these materials may be disordered organic semiconductors. Several studies since have claimed to show that melanins in the condensed solid-state are indeed semiconductors (9,10). However, it is by no means certain that the conductivity reported is electronic in nature. A clear idea of the basic structural unit is fundamental to developing a consistent model for condensed phase charge transport in such disordered organic systems.

Recently, steady state and time-resolved optical spectroscopies have shed some light on the photophysics and photochemistry of melanins. Additionally, such techniques have been used to probe electronic and molecular structure in order to address the vexing question of the basic structural unit. In particular, Nofsinger *et al*. have published several comprehensive steady state and dynamic studies (11,12). They report that sepia eumelanin (a good model for human eumelanin) has a low radiative relaxation quantum yield ($\phi_r = 3x10^{-3}$). Additionally, they have found that the time decay of the emission is non-exponential, and postulate that this may arise from a number of chemically distinct species. This proposition is supported by measurements on isolated mass fractions, which clearly show that radiative relaxation in sepia eumelanin is affected by aggregation. Their work lends support to the theory of an oligomeric basic structural unit, and from a photophysical perspective, suggests that non-radiative relaxation modes are the dominant energy dissipating pathways (13,7).

In general, there are relatively few reports of quantitative emission and excitation studies on melanins. Notable works include an early study by Kozikowski *et al*. (14), a detailed wavelength and concentration dependent study by Gallas and Eisner (15), and steady-state measurements on opioimelanins by Mosca *et al*. (16). The lack of quantitative data (radiative relaxation quantum yields in particular) is related to the inherent difficulties of performing optical spectroscopy on broad band absorbing macromolecules. Photoluminescence emission from melanins is heavily re-absorbed across a large range of wavelengths. This phenomenon, coupled with highly non-linear absorption and heavy pump beam attenuation, makes extracting quantitative information from emission or excitation data very difficult without appropriate re-normalisation. In general, the effects of probe beam attenuation and re-absorption in melanin studies are minimized by using low concentration samples. The aforementioned estimates of radiative relaxation quantum yield by Nofsinger *et al.* were made in this low concentration limit (11,12). Gallas and Eisner (15) report a correction method, but then fail to apply it to produce a meaningful quantum yield.

In this paper we report absolute values for radiative relaxation quantum yield for a synthetic eumelanin. These values were obtained by applying a re-normalisation procedure to account for probe beam attenuation and emission re-absorption in melanin samples and the fluorescein standards over a large range of concentrations. We have determined quantum yields as a function of excitation wavelength, and found a clear and systematic dependency. This work is motivated by a desire to gain a better understanding of melanin photophysics and photochemistry, and also, to address the fundamental question of the melanin basic structural unit. As such, it is part of a broader integrated program of quantum chemical, molecular and condensed phase (solid-state) studies aimed at elucidating the structure-property-function relationships of melanins.

## MATERIALS AND METHODS

*Sample Preparation.* Synthetic eumelanin (dopamelanin) derived from the non-enzymatic oxidation of tyrosine was purchased from Sigma Aldrich (Sydney, Australia), and used without further purification. Eumelanin solutions were prepared at a range of concentrations (0.001% to 0.005%) by weight macromolecule in high purity 18.2 MΩ MilliQ de-ionised water. To aid solubility, the pH of the solutions was adjusted using 0.01 M NaOH to ~11.5, and the solutions gently heated with stirring. Under such conditions, pale brown, apparently continuous eumelanin dispersions were produced. Fluorescein ($\phi_r$ =0.92 ± 0.02) was purchased from Sigma Aldrich (Sydney, Australia) and used without further purification to prepare standard solutions at ten different concentrations varying from $1.2 \times 10^{-4}$ % to $5 \times 10^{-6}$ % by weight in 0.1M NaOH solution (18.2 MΩ MilliQ de-ionised water). Fluorescein and eumelanin concentrations were chosen so as to maintain absorbance levels within the range of the absorption spectrometer.

*Absorption Spectrometry*. Absorption spectra between 200 and 800 nm were recorded for the synthetic eumelanin and fluorescein solutions using a Perkin Elmer (Melbourne, Australia) λ40 spectrophotometer. An integration of 2 nm, scan speed of 240 nm/min and slit width of 3 nm were used. Spectra were collected using a quartz 1 cm square cuvette. Solvent scans (obtained under identical conditions) were used for background correction.

*Photoluminescence Emission Spectrometry*. Photoluminescence emission spectra for the eumelanin and fluorescein solutions were recorded for all concentrations using a Jobin Yvon (Paris, France) FluoroMax 3 Fluorimeter. Emission scans were performed between 400 nm and 700 nm using excitation wavelengths of between 350 nm and 410 nm for the eumelanin samples, and 490 nm for the fluorescein samples. A band pass of 3 nm and an integration of 0.3 s were used. The PL spectra were corrected for attenuation of the probe beam and re-absorption of the emission according to the procedure outlined below. Background scans were performed under identical instrumental conditions using the relevant solvents. Spectra were automatically corrected to account for differences in pump beam power at different excitation wavelengths.

*Emission Re-normalisation*. All emission spectra were re-normalised to account for probe beam attenuation and emission re-absorption. The following procedure was used:

The measured PL emission intensity ($I_m(\lambda)$) at any particular excitation wavelength is related to the actual PL emission intensity ($I_c(\lambda)$) via the relationship:

$$I_c(\lambda) = I_m(\lambda) \cdot k(\lambda) - I_{bg}(\lambda) \tag{1}$$

where, ($I_{bg}(\lambda)$) is the background contribution to the measured emission intensity (solvent and impurity emission, and Raman scattering of the excitation by the solvent), and $k(\lambda)$ is a scaling factor defining the probe beam attenuation and emission re-absorption. If we assume that only emission arising from a small volume at the waist of the excitation beam is collected by the spectrometer detection system, then $k(\lambda)$ can be written as:

$$k(\lambda) = exp(\alpha_1 d_1 + \alpha_2(\lambda) d_2) \tag{2}$$

where, $\alpha_1$ is the absorption coefficient ($cm^{-1}$) at the excitation wavelength, $d_1$ (cm) is the effective path length responsible for attenuating the excitation beam, and $\alpha_2(\lambda)$ ($cm^{-1}$) and $d_2$ (cm) are respectively the wavelength dispersive absorption coefficient (over the emission range) and path length responsible for emission re-absorption. Eq. 1 also holds for PLE measurements, but in this case, the scaling factor $k(\lambda)$ is given by:

$$k(\lambda) = exp(\alpha_1(\lambda) d_1 + \alpha_2 d_2) \tag{3}$$

The only difference between Eq. 2 and Eq. 3 is that $\alpha_1(\lambda)$ is now dispersive in wavelength and $\alpha_2$ is the absorption coefficient at the detection wavelength. If the geometry of the measurement system is known, then the path lengths $d_1$ and $d_2$ can be found by inspection. For example, for a collimated excitation beam incident upon a square cross-section cuvette (sides of length $x$ cm) with collection at 90° with respect to excitation (and in the same horizontal plane), then:

$$d_1 = d_2 = \frac{x}{2} \tag{4}$$

However, if the geometry of the system is ill-defined, an estimate for the scaling factor can be found by analyzing the relative attenuation of the Raman scattered probe in the sample with respect to the solvent. This technique is useful in the case of melanin emission and excitation studies, since the relatively weak photoluminescence is the same order of magnitude as the signal resulting from Raman scattering of the probe from water (the solvent). In this case, Eqs. 2 and 3 (PL emission and excitation respectively) can be re-written as:

$$k(\lambda) = exp(\alpha_2(\lambda)d) \tag{5}$$

$$k(\lambda) = exp(\alpha_1(\lambda)d) \tag{6}$$

$$d = ln\left(\frac{I_{RB}}{I_R}\right) \cdot \frac{1}{\alpha_R} \tag{7}$$

where, $d$ is now an effective path length (reflecting both the pump beam attenuation and emission re-absorption), and $\alpha_1(\lambda)$ and $\alpha_2(\lambda)$ are as before. Additionally, $I_{RB}$ is the intensity of the Raman peak in the background, and $I_R$ is the attenuated Raman peak intensity in the sample. These values can be found by fitting Gaussian line shapes to the Raman and PL features in the emission spectra. The absorption coefficient at the Raman peak ($\alpha_R$) can be found from the absorbance spectra (Fig. 1). This method is particularly useful in PL thin film experiments where pump beam attenuation may be relatively weak, but emission re-absorption significant. In the data presented in this paper, all eumelanin and fluorescein emission spectra were re-normalised using Eqs. 1, 2 and 4, although we have shown that approximately similar results can be obtained using Eq. 5 to calculate $k(\lambda)$ (17). Absorption coefficients ($\alpha_1$ and $\alpha_2$) were determined from the absorbance spectra for each concentration. The errors associated with this re-normalisation procedure were estimated by considering the uncertainty in the scaling factor according to the equation:

$$\Delta k = k\left\{(\alpha_1 \cdot \Delta d_1)^2 + (\alpha_2(\lambda) \cdot \Delta d_2)^2\right\}^{1/2} \tag{8}$$

*Quantum Yield Calculations*. Radiative relaxation quantum yields at three excitation wavelengths (350 nm, 380 nm and 410 nm) were calculated for the synthetic eumelanin using a standard procedure (18). Plotting the integrated photoluminescence emission vs. absorbance for a range of concentrations allows the quantum yield ($\phi_{mel}$) to be determined according to the equation:

$$\phi_{mel} = \phi_s \left(\frac{g_{mel}}{g_s}\right) \cdot \left(\frac{n_{mel}}{n_s}\right)^2 \tag{9}$$

where, $\phi_s$ (0.92 ± 0.02 for fluorescein) is the known quantum yield of the standard, $n_{mel}$ and $n_s$ are the refractive indices of the melanin and standard solvents respectively (in this case both 1.33), and $g_{mel}$ and $g_s$ are the gradients of the integrated emission vs. absorbance plots. Given the dramatically different quantum yields of eumelanin and fluorescein, it was necessary to use neutral density filters (OD0.6 and OD1.0) to prevent detector saturation in the fluorescein emission measurements. These data were subsequently scaled prior to the re-normalisation process. The errors associated with the quantum yield measurements were calculated according to the equation (derived from Eq. 9):

$$\Delta\phi_{mel} = \phi_{mel} \cdot \left( \left( \frac{\Delta\phi_s}{\phi_s} \right)^2 + \left( \frac{\Delta g_{mel}}{g_{mel}} \right)^2 + \left( \frac{\Delta g_s}{g_s} \right)^2 \right)^{1/2} \qquad (10)$$

where, $\Delta g_{mel}$ and $\Delta g_s$ are the uncertainties associated with the gradients determined from the integrated emission vs. absorption plots. In this analysis, we have assumed $\Delta g_{mel}$, and $\Delta g_s$ and $\Delta\phi_s$ are uncorrelated.

## RESULTS & DISCUSSION

### Synthetic eumelanin absorption

It is well known that eumelanins in solution have a broad, monotonic absorption profile. In the absence of Rayleigh and Mie scattering, or any significant residual monomer or other contamination, the absorbance increases exponentially towards shorter wavelengths. This behavior is demonstrated in Fig. 1, which shows a series of spectra corresponding to concentrations of 0.001%, 0.0025% and 0.005% by weight synthetic eumelanin in pH11.5 de-ionised water. All spectra show a simple first order exponential dependency of absorption coefficient ($\alpha$) vs. wavelength. Indeed, a plot of $ln\alpha$ vs. $\lambda$ yields an approximate straight line (not shown). One could expect some contribution to attenuation from Rayleigh scattering. This would likely manifest itself at the low wavelength (high energy) end of the spectrum (below 350nm). Rayleigh scattering has a strong $\lambda^{-4}$ dependency (2), and hence a plot of $ln\alpha$ vs. $ln\lambda$ would give a gradient of approximately -4 in regions of the spectrum where the effect was significant. This was not observed in any of our samples. Mie scattering from un-dissolved aggregates is also a possible attenuation mechanism. This type of scattering is relatively wavelength independent, and is clearly visible as a broad feature in aggregated solutions at the concentrations of interest in our studies. None of the spectra showed such behavior. Hence, we conclude that our synthetic eumelanin was well solublised at pH 11.5, and that the attenuation shown in Fig. 1 is derived mainly from absorption. Fig. 2 shows a plot of absorption coefficient at 380 nm vs. concentration for the same three solutions. The relationship is clearly linear, and once again confirms that we have insignificant Mie scattering (this process would impose a non-linearity as a function of concentration). Our findings with respect to the absorption of eumelanin aqueous solutions at high pH agree with those reported by Nofsinger *et al*. (11), who found that under these conditions, no more than 15% of the attenuation could be attributed to scattering.

**Synthetic eumelanin photoluminescence emission**

Raw photoluminescence emission spectra (380 nm excitation wavelength) for synthetic eumelanin solutions at concentrations of 0.001%, 0.0025% and 0.005% by weight are shown in Fig. 3. There are several points to note in these raw spectra: a prominent Raman peak (Raman scattering of the excitation from the solvent), whose intensity is concentration dependent; the lack of correspondence between emission intensity and concentration; the variability in the wavelength of maximum emission; and the broad, single feature nature of the underlying emission. In theory, the emission intensity should scale linearly with concentration and the central maxima should occur at a constant wavelength. Significant attenuation of the excitation beam and re-absorption of the photoluminescence emission (both concentration dependent phenomena) are responsible for these departures from expected behavior. The application of Eqs. 1, 2 and 4 (with the absorption coefficient data from Fig. 1) corrects for these effects. Fig. 4 shows the re-normalised emission spectra resulting from this procedure.

We recover the expected intensity vs. concentration relationship, and the central maxima are now at the same wavelength. Fig. 5 confirms that the emission intensity is linearly dependent upon concentration. From this re-normalised data we can now see that the eumelanin emission is red shifted by ~100 nm relative to the excitation. The emission consists of a single, broad feature (not the double peaked feature reported by Gallas and Eisner (15)), similar to that seen by Nofsinger *et al.* (11,12) and Kozikowski *et al.* (14), with a full width at half maximum of ~ 130 nm.

The 0.0025% by weight sample was pumped at a number of additional excitation wavelengths (360 nm, 365 nm, 370 nm and 375nm). The spectra were re-normalised as above, and the data is plotted in Figs. 6a): wavelength and b): energy. There are two important points to note about these spectra: firstly, the relative intensities, widths and the positions of the emission maxima change with excitation wavelength (energy); and secondly, there appears to be a limiting value of the high wavelength (low energy) tail of the emission. These dependencies are summarized in Figs. 7a), b) and c). The plots are presented in energy units in order to remove the inherent non-linearity associated with wavelength. Given that we can discount any pump beam attenuation or emission re-absorption as causes for the phenomena seen in Fig. 7a), b) and c), we are led to the conclusion that this behavior is an inherent property of the eumelanin solution. One might argue that the reduction in photoluminescence emission at longer pump wavelengths is a function of the reduced absorption cross-section (Fig. 1). However, the quantum yield data (to follow) shows that the radiative emission efficiency is indeed excitation energy dependent.

**Radiative relaxation quantum yield**

The integrated photoluminescence emission vs. absorbance (at the excitation wavelength) plots for all fluorescein and synthetic eumelanin samples are shown in Figs. 8a) and b) respectively. Five concentrations (0.001% to 0.005%) were measured for the eumelanin, and ten ($1.2 \times 10^{-4}$ % to $5 \times 10^{-6}$ % by weight) were measured for fluorescein. The eumelanin samples were pumped at 380 nm and the fluorescein samples at 490 nm. Both plots show the effects of failure to correct for pump beam attenuation and emission re-absorption. Even the fluorescein samples at high absorbances (concentrations) show a marked deviation from the expected linearity. This effect is much more pronounced in the broad-band, heavily attenuating eumelanin solutions. Re-normalisation (as per the method outlined above) results in full recovery of linearity. Error bands were calculated according to Eq. 10. Similar curves were constructed for eumelanin samples pumped at 350 nm and 410 nm. Hence, absolute values of radiative relaxation quantum yield were determined for three excitation wavelengths according to Eq. 9. These values are shown in Fig. 9, and it is notable that all yields are between $5 \times 10^{-4}$ and $7 \times 10^{-4}$ - an order of magnitude lower than those reported by Nofsinger *et al.* (11,12). Additionally, there is a clear dependency of yield upon excitation energy: lower energy (longer wavelength) excitation leads to lower radiative emission. Nofsinger *et al.* have also observed changes in the emission spectra (positions of the peak maxima), and describe this as "*unusual since emission bands are generally insensitive in shape to the excitation wavelength*". However, they did not report excitation dependent yields, and assumed a constant value for all pump wavelengths.

We believe that our quantum yield values are an accurate reflection of the magnitude of radiative relaxation in synthetic eumelanins. Use of the re-normalisation procedure, and collection of data over a range of concentrations have improved the sensitivity and accuracy of our measurements versus previous estimates. It is interesting to note that the corrected integrated emission is linearly dependent upon concentration to a very good approximation. This would tend to indicate that other concentration dependent dynamic quenching mechanisms are not significant within this range. Additionally, the nature of the emission dependence upon excitation wavelength suggests that we are pumping chemically distinct species each with different fundamental HOMO-LUMO gaps. The species with the lowest energy gap (one could speculate that it was the largest oligomer capable of radiative emission) represents the “limiting” excitation energy for the ensemble (we see this as a clear low energy / high wavelength static tail). These species may be different sized DHI/DHICA oligomers, or oligomers containing a variety of DHI/DHICA tautomers (19). This picture fits with the broad band absorbance and the clear asymmetry between excitation and emission characteristics. Although our data is consistent with the oligomeric nano-aggregate theory for the basic structural unit, it does not completely preclude the co-existence of oligomers (the emissive species) and large heteropolymers (the non-emissive species) which serve to statically quench.

In conclusion, we have determined radiative relaxation quantum yields for synthetic eumelanin. A re-normalisation procedure was used to correct for pump beam attenuation and heavy re-absorption of the emission. These are particular issues with broad band absorbing materials such as melanins. By applying the procedure, we were able to obtain data over a range of concentrations and for several different pump wavelengths. Our results confirm that eumelanin is capable of dissipating >99.9% of the absorbed UV and visible radiation non-radiatively. Additionally, we have found a clear and systematic dependency of emission upon excitation energy. The position, width and intensity of the emission maxima, as well as the quantum yield, vary as a function of pump energy. Our data is consistent with emission arising from a number of chemically distinct species of different HOMO-LUMO gaps, and lends support to the argument that the basic structural unit of eumelanin is an oligomeric nano-aggregate rather than an extended heteropolymer. Further spectroscopic studies, Density Functional Theory quantum chemical simulations, and condensed phase measurements are currently in progress to shed further light upon this fundamental question.

*Acknowledgments*- This work has been supported in part by the Australian Research Council, the UQ Centre for Biophotonics and Laser Science, and the University of Queensland (RIF scheme). Our thanks go to Prof. Ross McKenzie and Dr. Ben Powell for stimulating discussion with respect to the basic structural unit question, and Prof. Tad Sarna and Prof. John Simon concerning the spectroscopic findings.

## REFERENCES

1. Prota, G. (1992) *Melanins and Melanogenesis*. Academic Press, San Diego, CA.
2. Wolbarsht, M. L., A. W. Walsh and G. George (1981) Melanin, a unique biological absorber. *Appl. Opt*. **20**, 2184-2186.
3. Hill, H. Z. (1995) Is melanin photoprotective or is it photosensitizing? *In Melanin: Its role in human photoprotection* (Edited by L. Zeise, M. Chedekel and T. Fitzpatrick), pp. 81-91. Valdenmar Press, Overland Park, KS.
4. Stark, K. B., J. M. Gallas, G. W. Zajac, M. Eisner and J. T. Golab (2003) Spectroscopic study and simulation from recent structural models for eumelanin: I monomers, dimers. *J. Phys. Chem. B* **107**, 3061-3067.
5. Zajac, Z. W., J. M. Gallas, J. Cheng, M. Eisner, S. C. Moss and A. E. Alvarado-Swaisgood (1994) The fundamental unit of synthetic melanin: a verification by tunneling microscopy and X-ray scattering results. *Biochim. Biophys. Acta* **1199**, 271-278.
6. Ito, S. (1986) Reexamination of the structure of eumelanin. *Biochim. Biophys. Acta* **883**, 155-161.
7. Clancy, C. M. R., J. B. Nofsinger, R. K. Hanks and J. D. Simon (2000) A hierarchical self-assembly of eumelanin*. J. Phys. Chem. B* **104**, 7871-7873.
8. McGinness, J., P. Corry and P. Proctor (1974) Amorphous semiconductor switching in melanins. *Science* **183**, 853-855.
9. Crippa, P. R., V. Cristofoletti and N. Romeo (1978) A band model for melanin deduced from optical absorption and photoconductivity experiments. *Biochim. Biophys. Acta* **538**, 164-170.
10. Jastrzebska, M. M., H. Isotalo, J. Paloheimo, H. Stubb and B. Pilawa (1996) Effect of

$Cu^{2+}$-ions on semiconductor properties of synthetic DOPA melanin polymer. *J. Biomater. Sci. Polymer Edn.* **7**, 781-793.

11. Nofsinger, J. B. and J. D. Simon (2001) Radiative relaxation of sepia eumelanin is affected by aggregation. *Photochem. Photobiol.* **74**, 31-37.
12. Nofsinger, J. B., T. Ye and J. D. Simon (2001) Unltrafast nonradiative relaxation dynamics of eumelanin. *J. Phys. Chem. B* **105**, 2864-2866.
13. Nofsinger, J. B., S. E. Forest and J. D. Simon (1999) Explanation for the disparity among absorption and action spectra for eumelanin. *J. Phys. Chem. B* **103**, 11428-11432.
14. Kozikowski, S. D., L. J. Wolfram and R. R. Alfano (1984) Fluorescence spectroscopy of eumelanins. *IEEE J. Quant. Electron.* **QE-20**, 1379-1382.
15. Gallas, J. M. and M. Eisner (1987) Fluorescence of melanin – dependence upon excitation wavelength and concentration. *Photochem. Photobiol.* **45**, 595-600.
16. Mosca, L., C. DeMarco, M. Fontana and M. A. Rosei (1999) Fluorescence properties of melanins from opioid peptides. *Arch. Biochim. Biophys.* **371**, 63-69.
17. Riesz, J., J. Gilmore and P. Meredith (2003) Quantitative photoluminescence of broad band absorbing melanins: a procedure to correct for inner filter and re-absorption effects. *submitted Spectrochimica Acta Part A.*
18. Lakowicz, J. R. (1999) *Principles of Fluorescence Spectroscopy*. 2nd Ed. Kluwer Academic/Plenum Publishers, New York.
19. Il'ichev, Y. V. and J. D. Simon (2003) Building blocks of eumelanin: relative stability and excitation energies of tautomers of 5,6-dihydroxyindole and 5,6-indolequinone. *J. Phys. Chem. B* **107**, 7162-7171.

## FIGURE CAPTIONS

**Figure 1** Absorption spectra (absorption coefficient vs. wavelength) for three synthetic eumelanin solutions: 0.005% (dotted line), 0.0025% (dashed line) and 0.001% (solid line) by weight concentration.

**Figure 2** Absorption coefficient at 380nm vs. concentration for the three synthetic eumelanin solutions in Fig. 1. Concentration errors were estimated to be $1 \times 10^{-4}$% by weight.

**Figure 3** Raw PL emission spectra (pumped at 380 nm) for three synthetic eumelanin solutions: 0.005% (dotted line), 0.0025% (dashed line) and 0.001% (solid line) by weight concentration, and solvent background (dot-dash line).

**Figure 4** Re-normalised PL emission spectra (pumped at 380 nm) for three synthetic eumelanin solutions: 0.005% (dotted line), 0.0025% (dashed line) and 0.001% (solid line) by weight concentration.

**Figure 5** Re-normalised PL emission peak intensity vs. concentration for three synthetic eumelanin solutions: 0.005%, 0.0025% and 0.001% by weight concentration. The samples were pumped at 380 nm. Intensity errors were calculated according to the accuracy of the re-normalisation procedure (Eq. 8).

**Figure 6** Re-normalised PL emission spectra for a 0.0025% by weight synthetic eumelanin solution: a) plotted vs. wavelength, and b) plotted vs. energy for five pump wavelengths: 360 nm (solid line) to 380 nm (inner dashed line) in 5 nm increments.

**Figure 7** Re-normalised PL emission peak: a) full width at half maxima, b) position and c) intensity vs. excitation energy for a 0.0025% synthetic eumelanin sample pumped at 360 nm, 365 nm, 370 nm, 375 nm and 380 nm. Errors were derived from the re-normalisation procedure (Eq. 8), and subsequent Gaussian fitting.

**Figure 8a)** Integrated PL emission vs. absorption coefficient at the excitation wavelength (490 nm) for ten fluorescein solutions ($1.2 \times 10^{-4}$ % to $5 \times 10^{-6}$ % by weight): raw data (open circles) and re-normalised to account for pump attenuation and emission re-absorption (solid line). Errors were derived from the re-normalisation procedure (Eq. 8).

**Figure 8b)** Integrated PL emission vs. absorption coefficient at the excitation wavelength (380 nm) for five melanin solutions (0.001% to 0.005%): raw data (open circles) and re-normalised to account for pump attenuation and emission re-absorption (solid line). Errors were derived from the re-normalisation procedure (Eq. 8).

**Figure 9** Radiative relaxation quantum yield for synthetic eumelanin pumped at three wavelengths (350 nm, 380 nm and 410 nm). Errors were calculated according to Eq. 10. The solid line is a linear fit which is meant only as a guide to the eye.

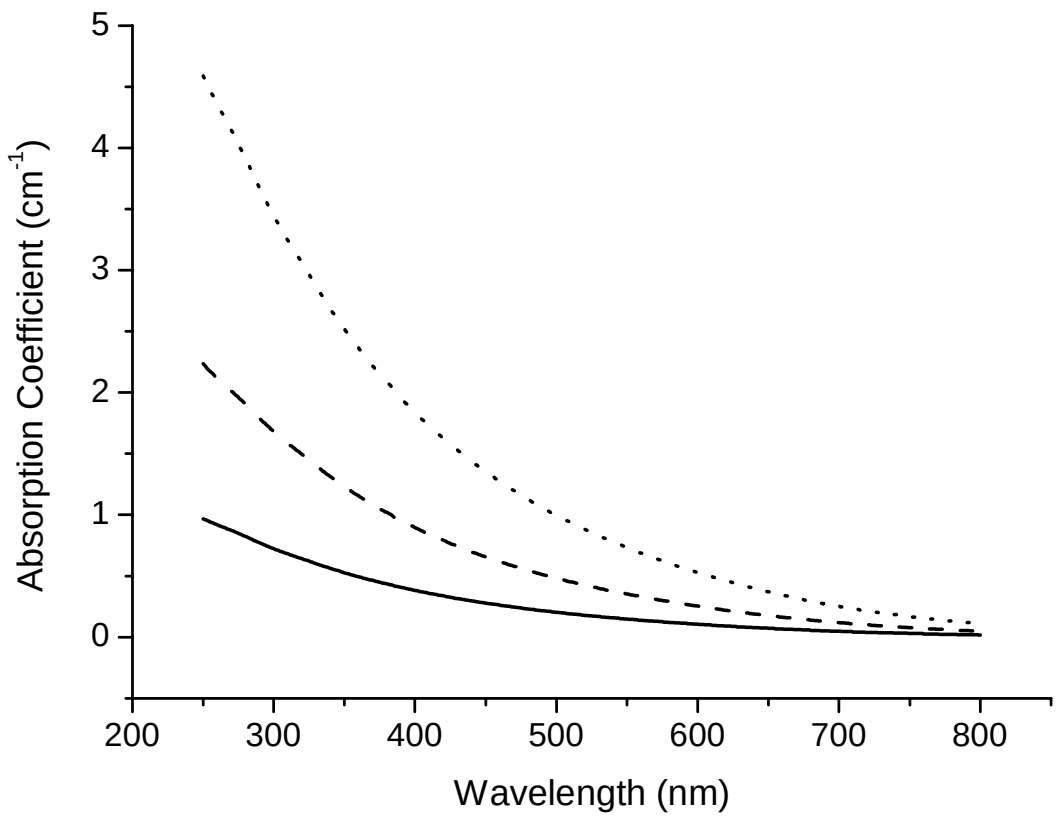

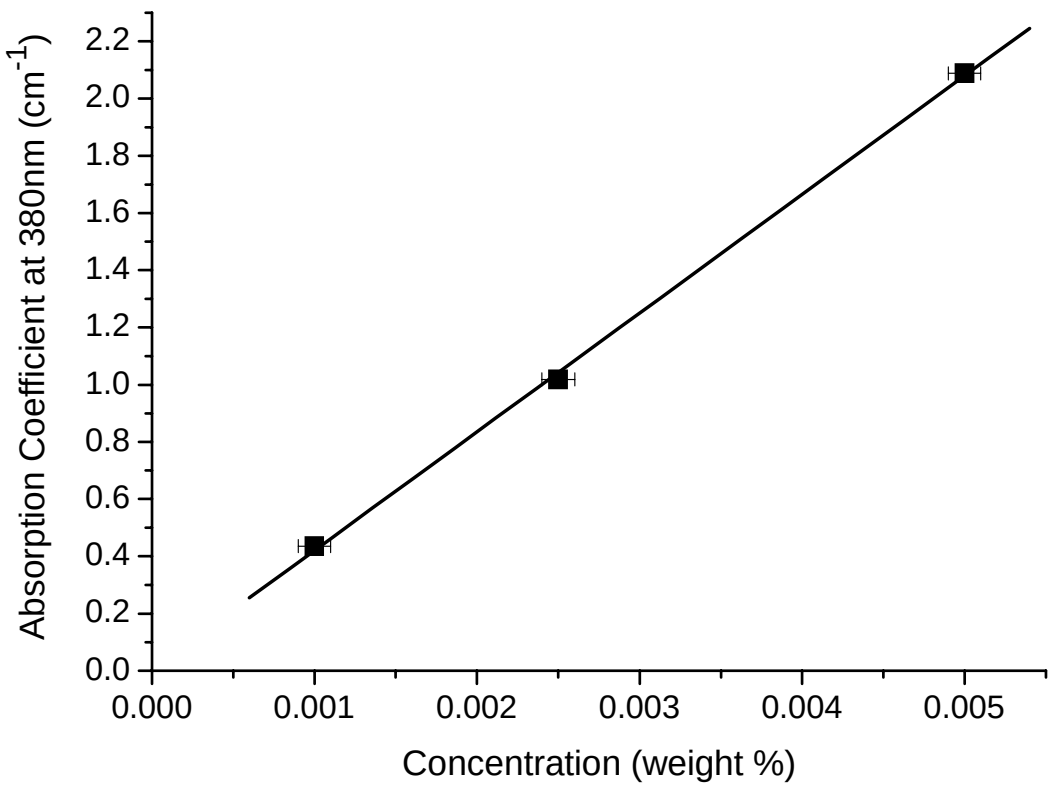

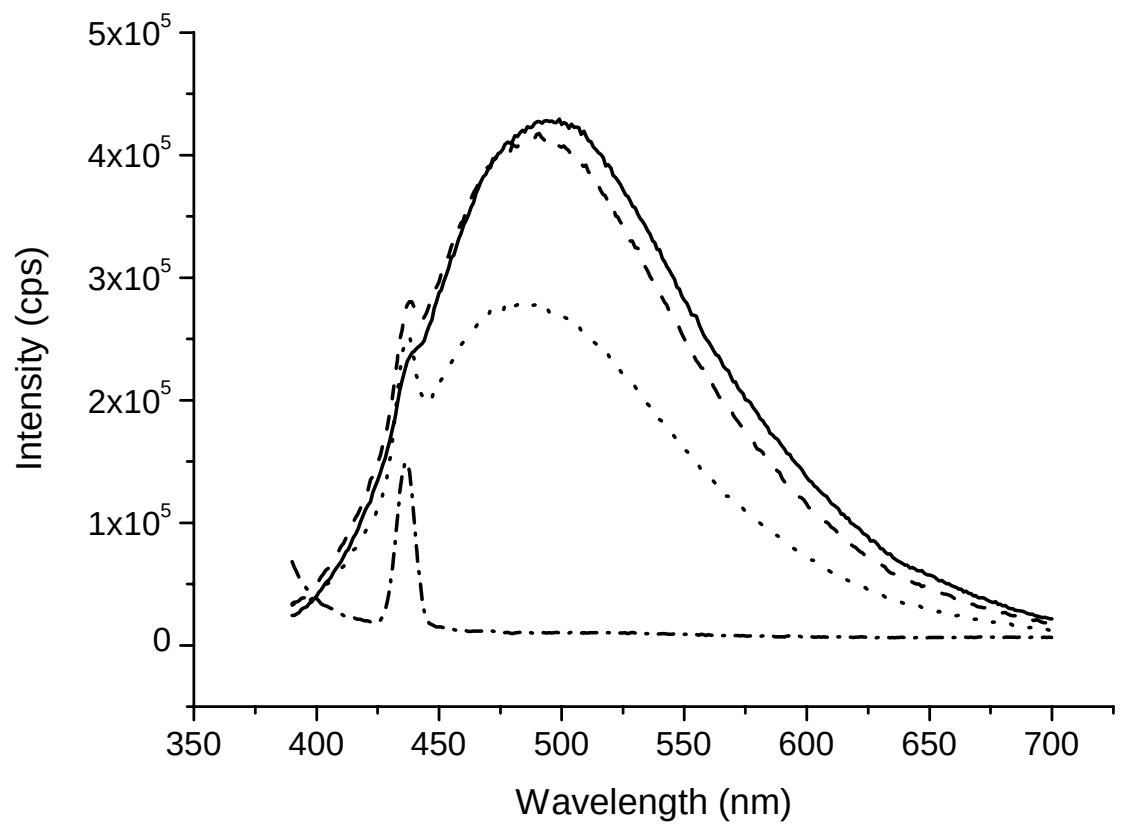

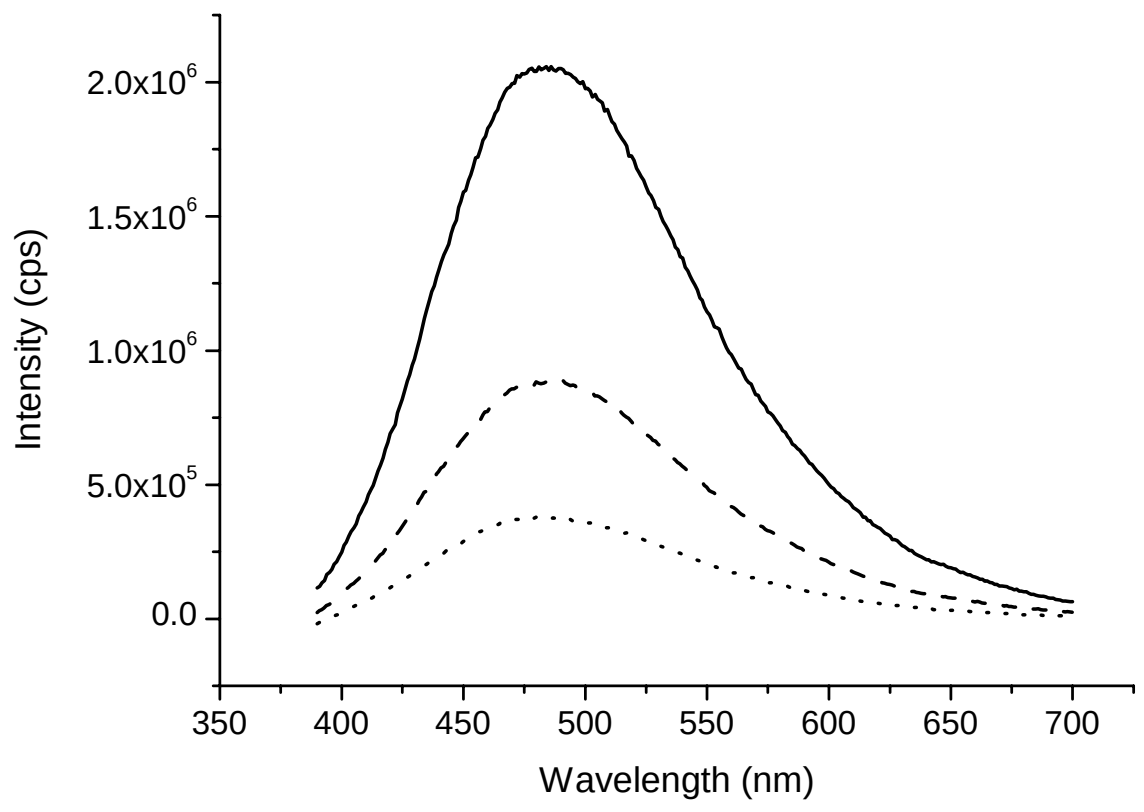

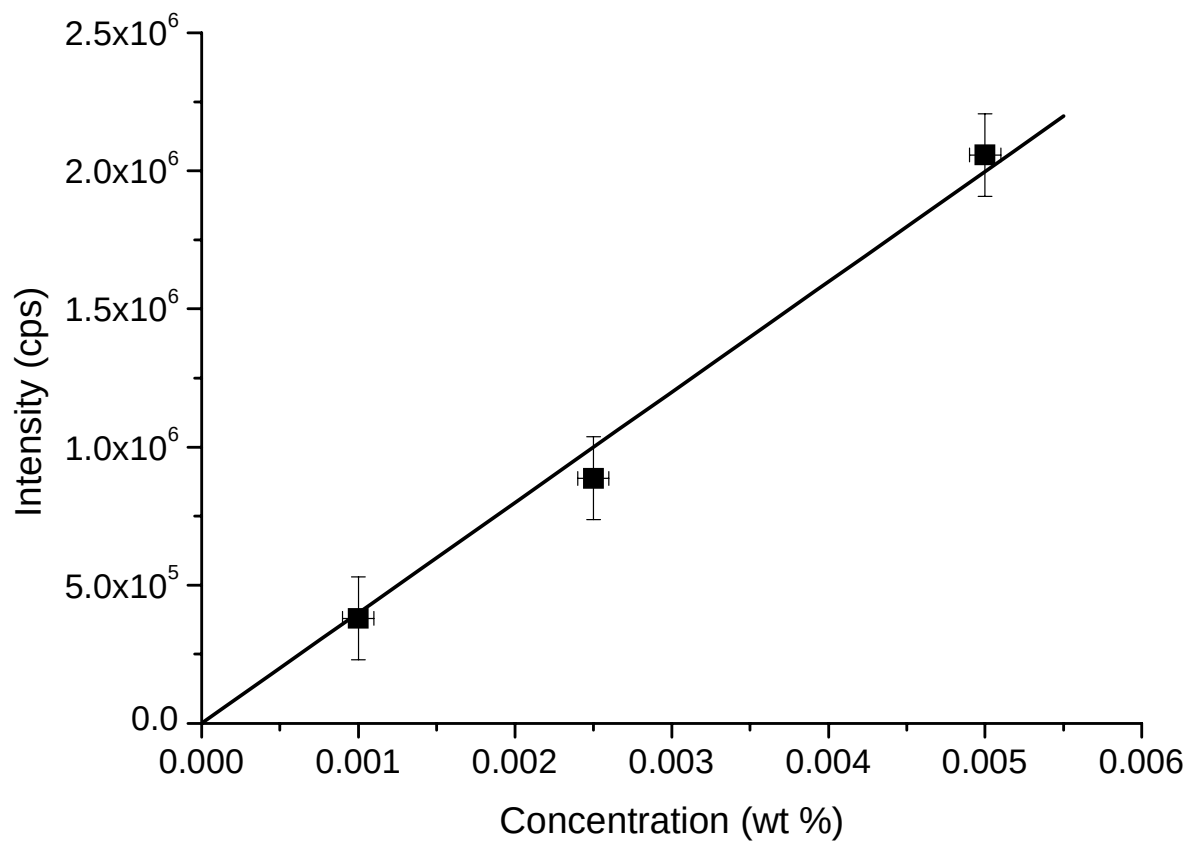

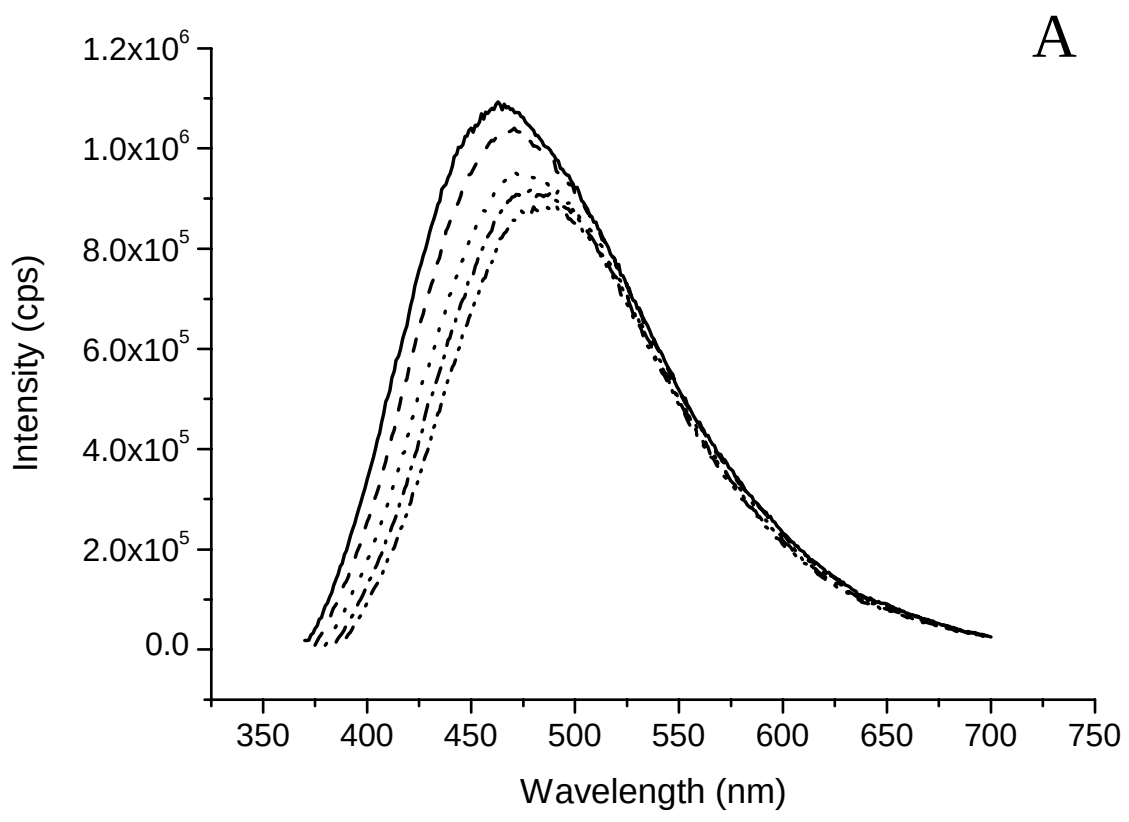
A
1.2x10^6
1.0x10^6
8.0x10^5
6.0x10^5
4.0x10^5
2.0x10^5
0.0
Intensity (cps)
350 400 450 500 550 600 650 700 750
Wavelength (nm)

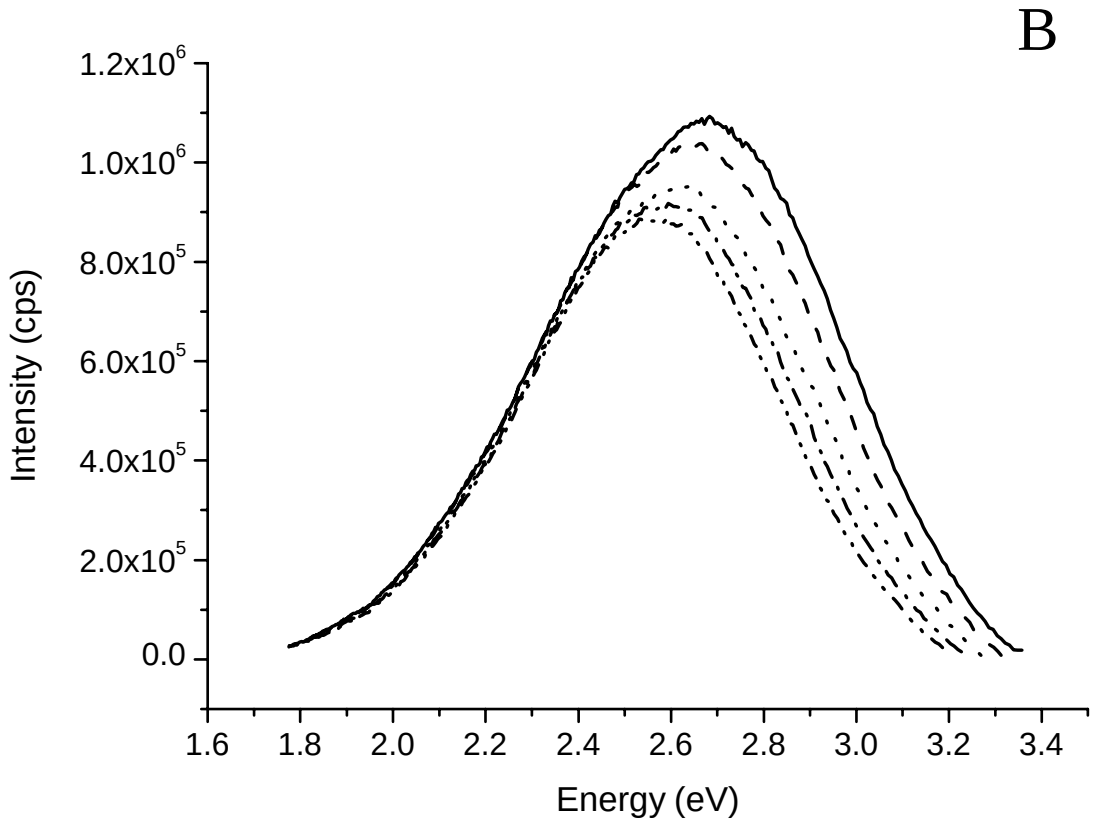
B
1.2x10^6
1.0x10^6
8.0x10^5
6.0x10^5
4.0x10^5
2.0x10^5
0.0
Intensity (cps)
1.6 1.8 2.0 2.2 2.4 2.6 2.8 3.0 3.2 3.4
Energy (eV)

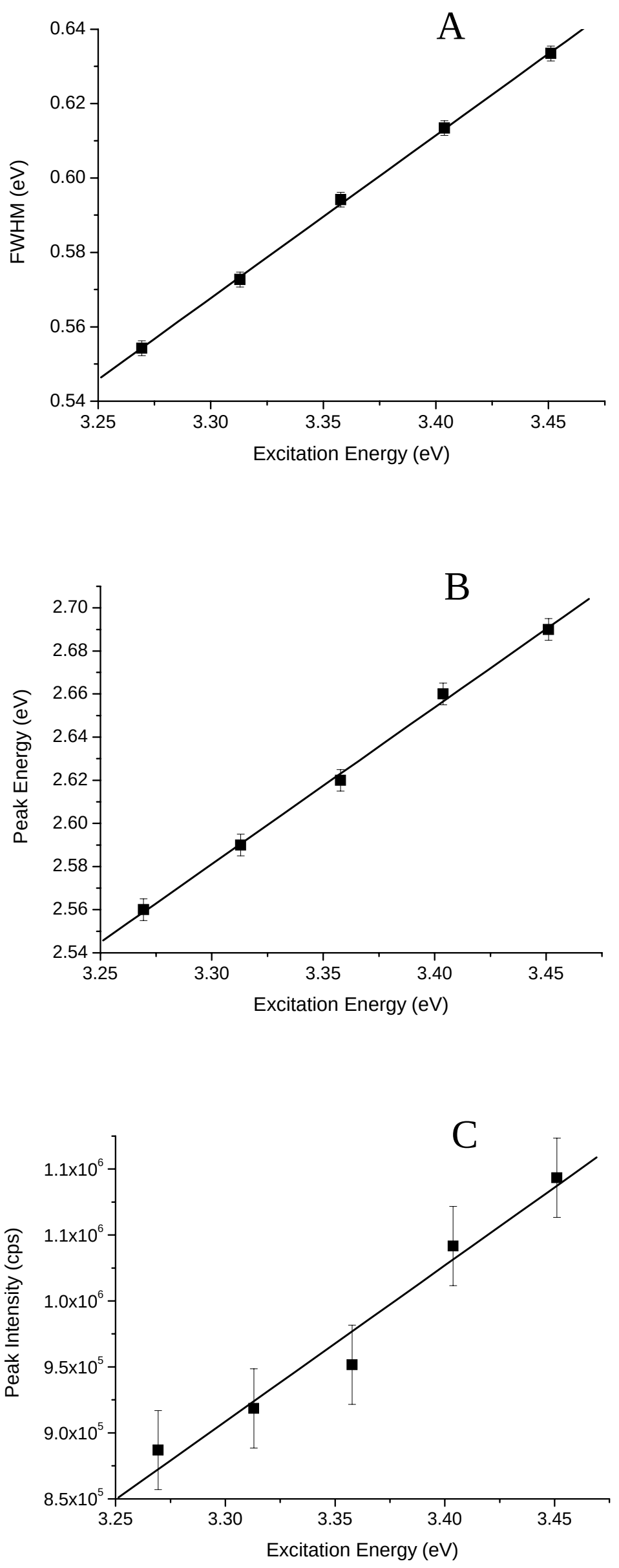

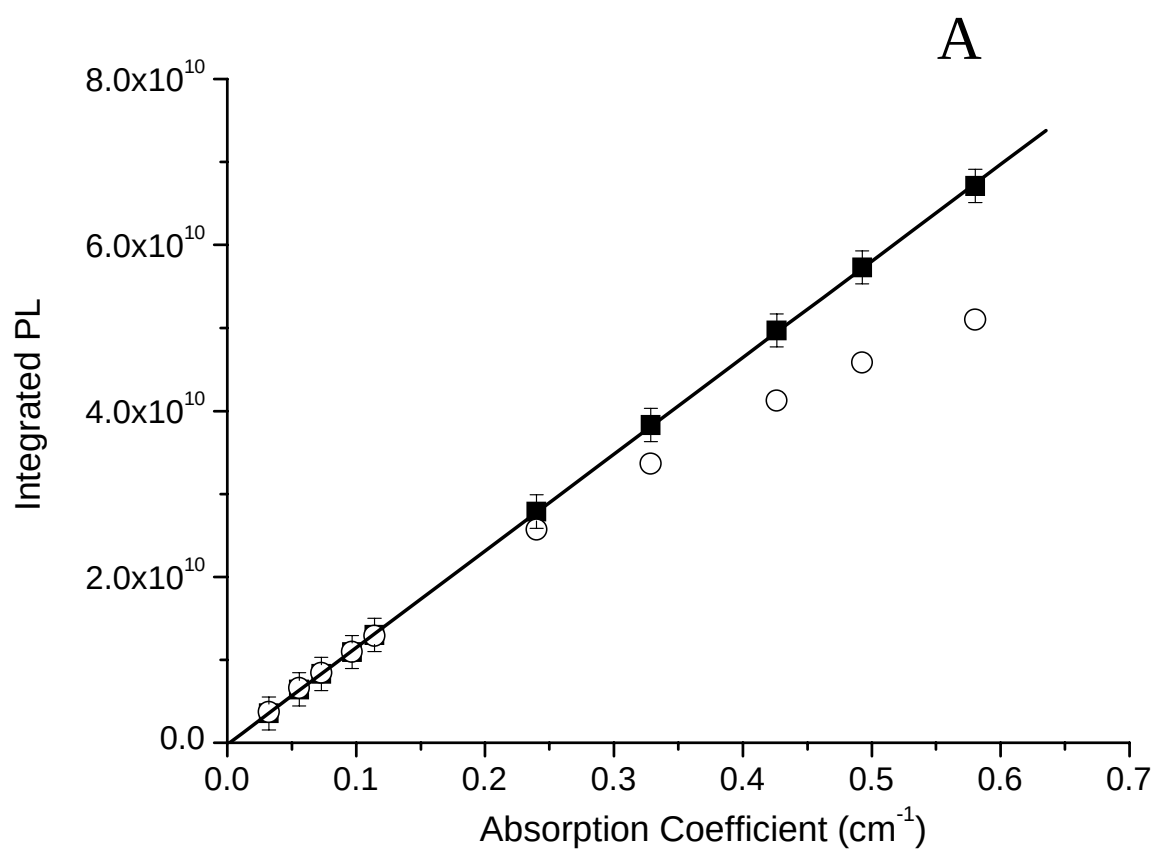

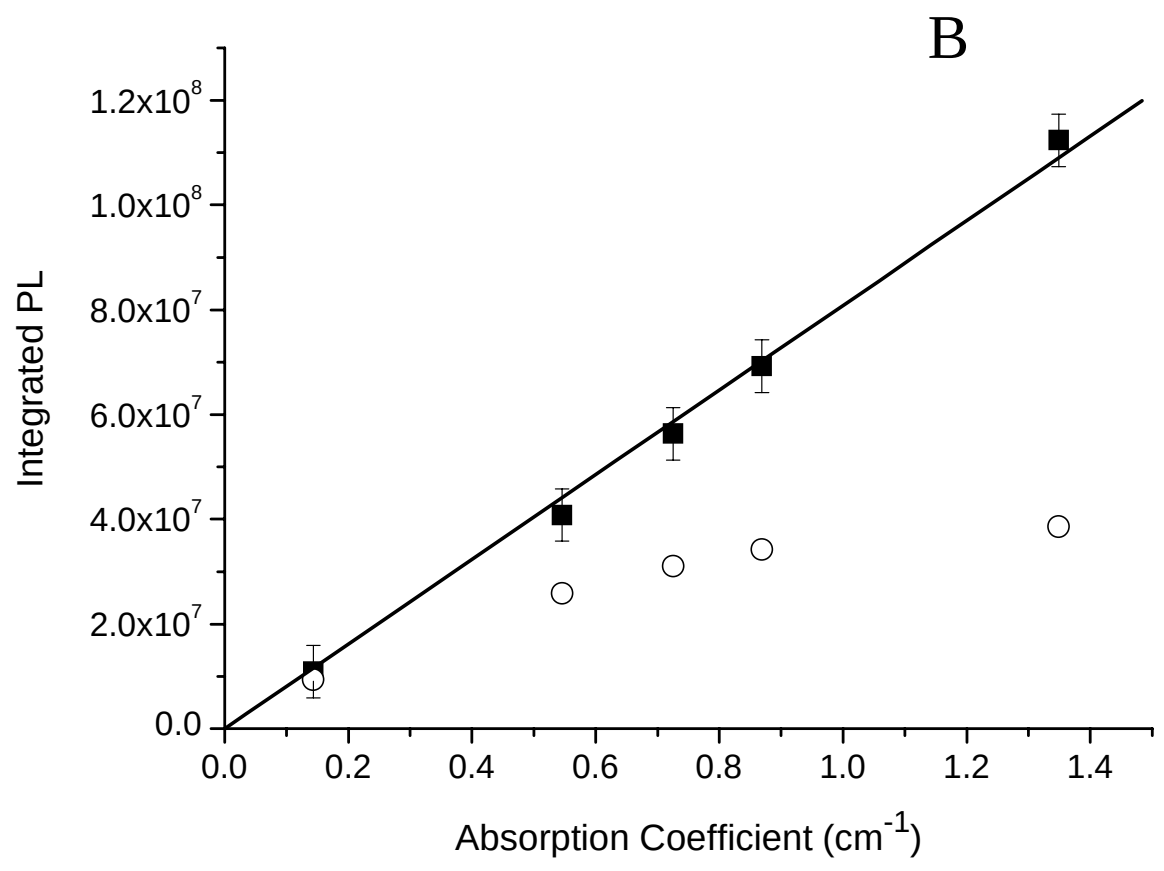

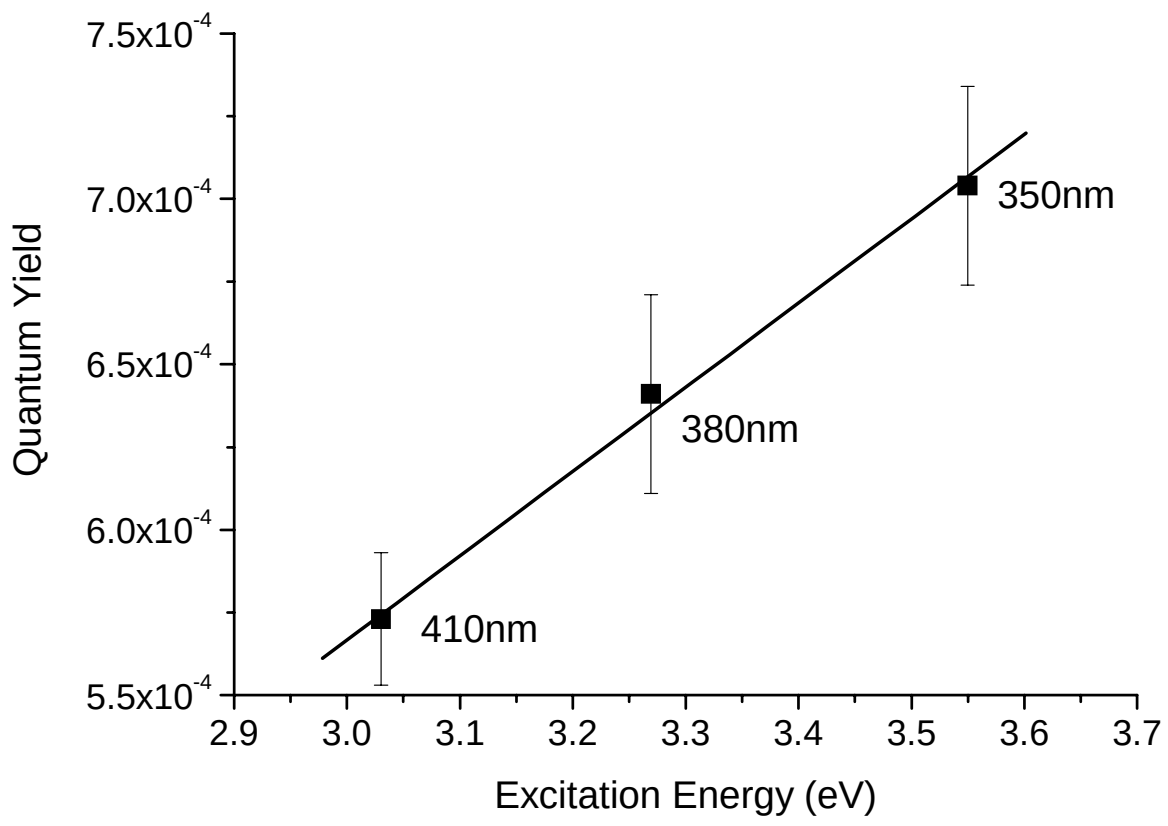